\documentclass{blois}
\pdfoutput=1 
\usepackage{jinstpub} 
\notoc

\usepackage[colorinlistoftodos]{todonotes}
\usepackage{HGTD_TDR-defs}
\usepackage{multirow}
\usepackage{rotating}
\usepackage{mathrsfs}
\usepackage{booktabs}
\usepackage{longtable}
\usepackage{placeins}
\usepackage{amssymb}
\usepackage{upgreek}
\usepackage{svg}
\usepackage{comment}
\usepackage{pdfpages}
\usepackage{siunitx}
\usepackage{lineno}
\usepackage{float}
\usepackage{lineno}
\usepackage{comment}
\usepackage{flafter}
\usepackage{subcaption}

\newcommand{\tmax}{{t_\text{max}}}
\newcommand{\pmax}{{p_\text{max}}}

\title{\boldmath Synchrotron light source focused X-ray detection with LGADs, AC-LGADs and TI-LGADs}

\abstract{
The response of Low Gain Avalanche Diodes (LGADs), a type of thin silicon detector with internal gain, to X-rays of energies between 6-16~keV was characterized at the Stanford Synchrotron Radiation Lightsource (SSRL). 
The utilized beamline at SSRL was 7-2, with a nominal beam size of 30~$\mu$m, repetition rate of 500~MHz, and with an energy dispersion $\Delta E/E$ of $10^{-4}$. 
Multi-channel LGADs, AC-LGADs, and TI-LGADs of different thicknesses and gain layer configurations from Hamamatsu Photonics (HPK) and Fondazione Bruno Kessler (FBK) were tested.
The sensors were read out with a discrete component board and digitized with a fast oscilloscope or a CAEN fast digitizer.
The devices' energy response, energy resolution, and time resolution were measured as a function of X-ray energy and position. The charge collection and multiplication mechanism were simulated using TCAD Sentaurus, and the results were compared with the collected data.
}

\author[1]{A.~Molnar}
\author[1]{M.~Davis}
\author[1]{G.~Oregan}
\author[1]{S.~Beringer}
\author[2]{Y.~Zhao}
\author[1]{S.M.~Mazza\footnote{Corresponding author}}
\author[1]{A.~Tiernan}
\author[3]{J.~Ott}
\author[1]{H.~F.-W.~Sadrozinski}
\author[1]{A.~Seiden}
\author[1]{B.~Schumm}
\author[1]{F.~McKinney-Martinez}
\author[4]{A.~Bisht}
\author[4]{M.~Centis-Vignali}
\author[4]{G.~Paternoster }
\author[4]{M.~Boscardin}

\affiliation[1]{SCIPP, University of California Santa Cruz, 1156 High Street, Santa Cruz (CA), US}
\affiliation[2]{Physics Department, Carleton University, 1125 Colonel By Drive, Ottawa, Ontario, K1S 5B6, Canada}
\affiliation[4]{Fondazione Bruno Kessler}
\affiliation[3]{Department of Electrical and Computer Engineering, University of Hawaii at Manoa, 2540 Dole Street, Honolulu HI-96822, USA}

\emailAdd{simazza@ucsc.edu}


\arxivnumber{2504.18638} 

\begin{document}







\maketitle
\flushbottom

\section{Introduction}
\label{sec:intro}

LGADs are silicon sensors with moderate internal gain and tens of picoseconds time resolution~\cite{bib:LGAD,bib:UFSD300umTB}. 
They comprise a low-doped region called 'bulk', typically 20-50~\si{\micro\meter} thick for fast timing applications (e.g. 50~\si{\micro\meter} for the ATLAS/CMS timing layers), and a highly doped thin region, a few \si{\micro\meter} from the charge collection electrode, called the 'gain layer'. Thin LGADs have a fast rise time, exceptional time resolution, and short full charge collection time.

The response of a number of LGAD detectors to X-rays of energies 6-16~\si{\kilo\electronvolt} (with a  $\Delta E/E$  of $10^{-4}$) was characterized at the Stanford Synchrotron Radiation Lightsource (SSRL)~\cite{SSRL} at the focused beamline 7-2~\cite{SSRL112}. The nominal beam size is \SI{30}{\micro\meter} and has a repetition rate of about \SI{500}{\mega\hertz}. 
LGADs with thicknesses of 50~$\mu$m, 100~$\mu$m, and 150~$\mu$m from Fondazione Bruno Kessler (FBK) and Hamamatsu Photonics (HPK) were used for this study.
The 50~$\mu$m-thin LGADs easily resolved in time the repetition rate of the beam line, while thicker LGADs suffer from pulse overlapping.
    The energy and time resolution of the tested LGADs were measured as a function of X-ray energy, device type, applied bias voltage, and gain. TCAD Sentaurus was used to understand the charge collection mechanism for X-ray interaction at different depths of the tested LGAD.

A previous study~\cite{Mazza:2023col} was performed at the SSRL beamline 11-2~\cite{SSRL112} with a non-focused beam but with a larger energy range (5-70 \si{\kilo\electronvolt}).
The effort to introduce the LGAD technology to the photon physics community is being pursued by several groups across the HEP community~\cite{iworid1,iworid2,GALLOWAY20195}.
The internal gain of the LGADs multiplies the charge deposited by the X-rays to boost the signal-to-noise ratio, allowing the detection of low-energy X-rays, while simultaneously enabling the rapid signal collection, due to the reduced thickness of these devices, for an ultra-high frame-rate response. 

\section{Devices tested}
\label{sec:devices}

\begin{table}[H]
    \centering
    \begin{tabular}{|l|c|c|c|c|c|}
        \hline
         Device & Fab & BV & Thickness & LGAD type & Geometry \\
         \hline
         FBK 'space' W1 & FBK & \SI{250}{\volt} & \SI{55}{\micro\meter} & DC-LGAD & single pad 3x3~\si{\milli\meter\squared}\\
         FBK 'space' W8 & FBK & \SI{400}{\volt} & \SI{100}{\micro\meter} & DC-LGAD & single pad 3x3~\si{\milli\meter\squared}\\
         FBK 'space' W11 & FBK & \SI{600}{\volt} & \SI{150}{\micro\meter} & DC-LGAD & single pad 3x3~\si{\milli\meter\squared}\\
         \hline
         FBK RSD2 various & FBK & \SI{350}{\volt} & \SI{50}{\micro\meter} & AC-LGAD & 4x4 array, 2x2~\si{\milli\meter\squared} \\
         \hline
         HPK-1 AC W2& HPK & \SI{200}{\volt} & \SI{50}{\micro\meter} & AC-LGAD & 500~$\mu$m pitch strip (5~\si{\milli\meter}) \\
         \hline
         FBK TI-LGAD strip & FBK & \SI{350}{\volt} & \SI{50}{\micro\meter} & TI-LGAD & 100~$\mu$m pitch strip (3~\si{\milli\meter})\\
         \hline
    \end{tabular}
    \caption{List of tested FBK and HPK LGADs. Fab. is the facility that produced the sensor. BV is the breakdown voltage.}
    \label{tab:LGADs}
\end{table}

\begin{figure}[H]
    \centering
    \begin{subfigure}{0.1\textwidth} 
    \includegraphics[width=\textwidth]
        {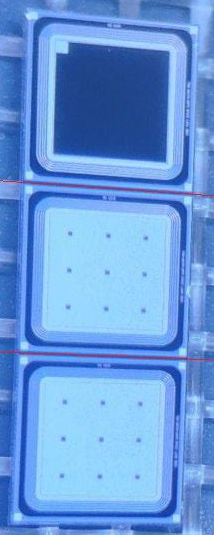}
            \caption{}
    \end{subfigure}
    \begin{subfigure}{0.15\textwidth} 
    \includegraphics[width=\textwidth]
        {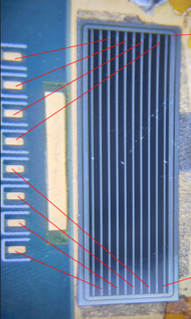}
            \caption{}
    \end{subfigure}
    \centering
    \begin{subfigure}{0.22\textwidth} 
    \includegraphics[width=\textwidth]
        {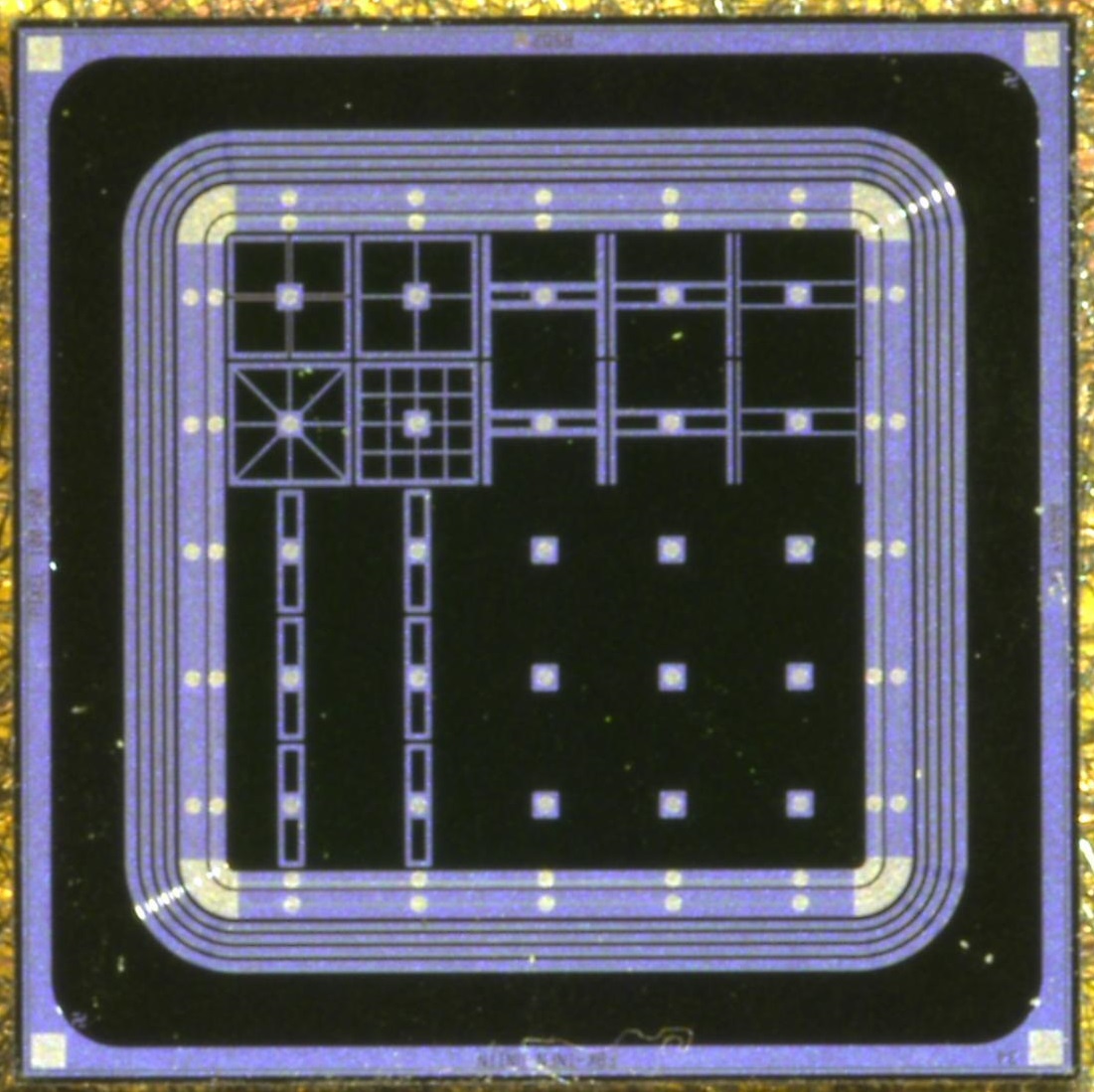}
            \caption{}
    \end{subfigure}
    \centering
    \begin{subfigure}{0.22\textwidth} 
    \includegraphics[width=\textwidth]{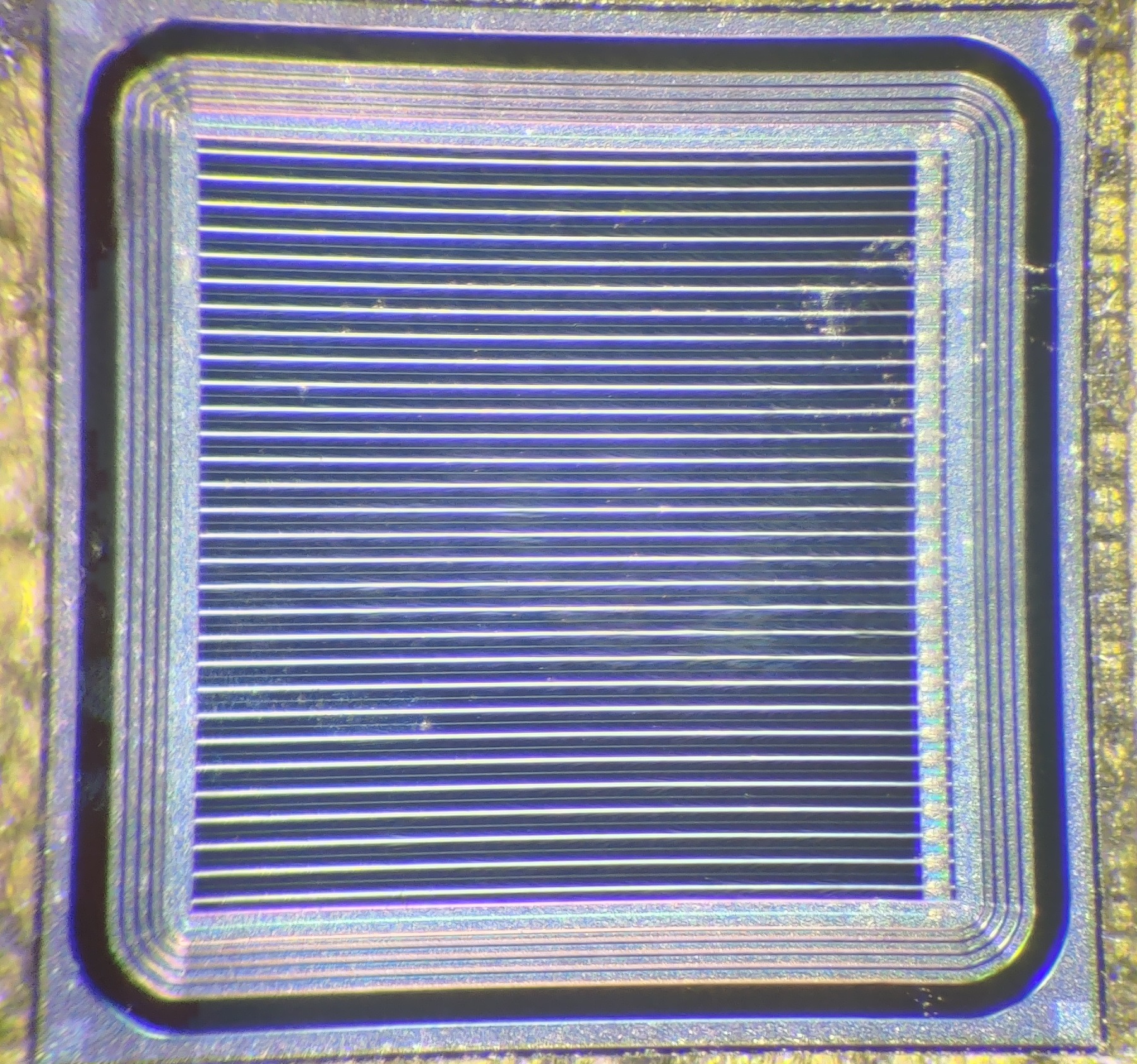}
            \caption{}
    \end{subfigure}

\caption{(a) FBK space single pads, (b) HPK strip detector, (c) FBK RSD2 various pad geometry, (d) FBK TI-LGAD strip}
\label{fig:LGADS}

\end{figure}

A full list of tested prototypes is shown in Tab.~\ref{tab:LGADs}.
A set of FBK prototypes produced for space applications was tested~\cite{Bisht:2024spa}, seen in Fig.~\ref{fig:LGADS} (a). 
These sensors are standard DC-LGADs with the same 3x3~mm single pad structure (in dies of 3 single pads); the gain layer is slightly different between devices.
The tested devices had \SI{55}{\micro\meter} (for simplicity referred to as \SI{50}{\micro\meter} in the following text), \SI{100}{\micro\meter}, and \SI{150}{\micro\meter} of active thickness.
The focused beam was positioned in the center of the detector, and the energy resolution, linearity, and time resolution were probed as a function of applied bias voltage and X-ray energy.
This production was selected because thicker LGADs are of interest for X-ray detection to increase absorption efficiency. 
Furthermore, the deeper depositions of these thicker sensors shed light on the charge transport, avalanche process, and gain suppression. 

Higher granularity LGADs, such as AC-LGADs and TI-LGADs, were tested with the focused X-ray beam. These sensors are particularly interesting for beam and image reconstruction at high frame rates.
The response of a pixel AC-LGAD from the FBK RSD2 production~\cite{Mandurrino:2021cuq} (Fig.~\ref{fig:LGADS} (c)) and strip AC-LGADs from the HPK ePIC 2023 production~\cite{Stage:2025jre} (Fig.~\ref{fig:LGADS} (b)) were tested as a function of position on the detector.
A strip TI-LGAD from the FBK RD50 production~\cite{Bisht:2022bvn} (Fig.~\ref{fig:LGADS} (d)) was also characterized as a function of position. 

\section{Experimental setup at SSRL}
\label{sec:setup}
\begin{figure}[H]
    \centering
    \begin{subfigure}{0.4\textwidth}  
    \includegraphics[width=\textwidth]
        {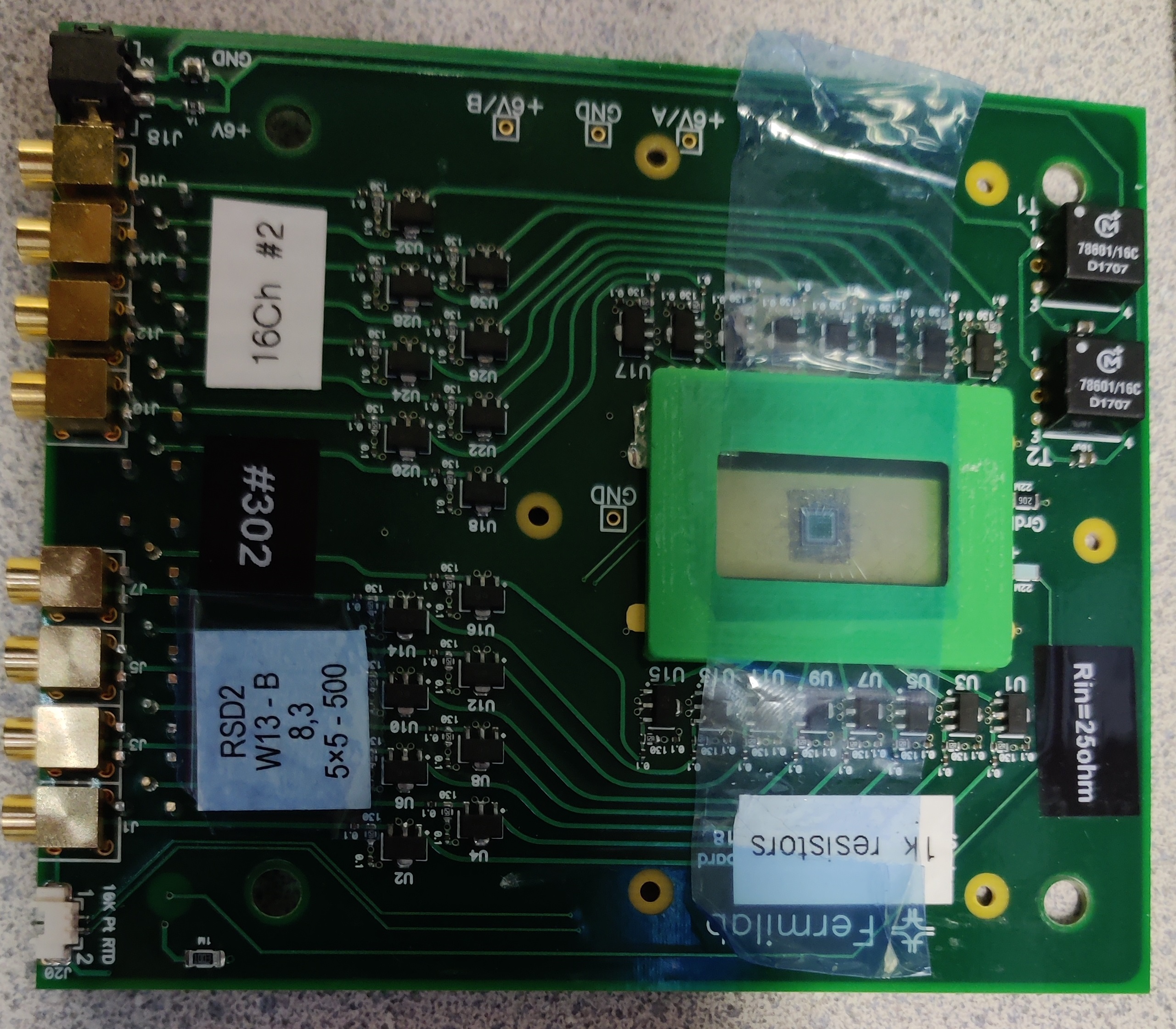}
        \caption{}
    \end{subfigure}
    \begin{subfigure}{0.47\textwidth}  
    \includegraphics[width=\textwidth]
        {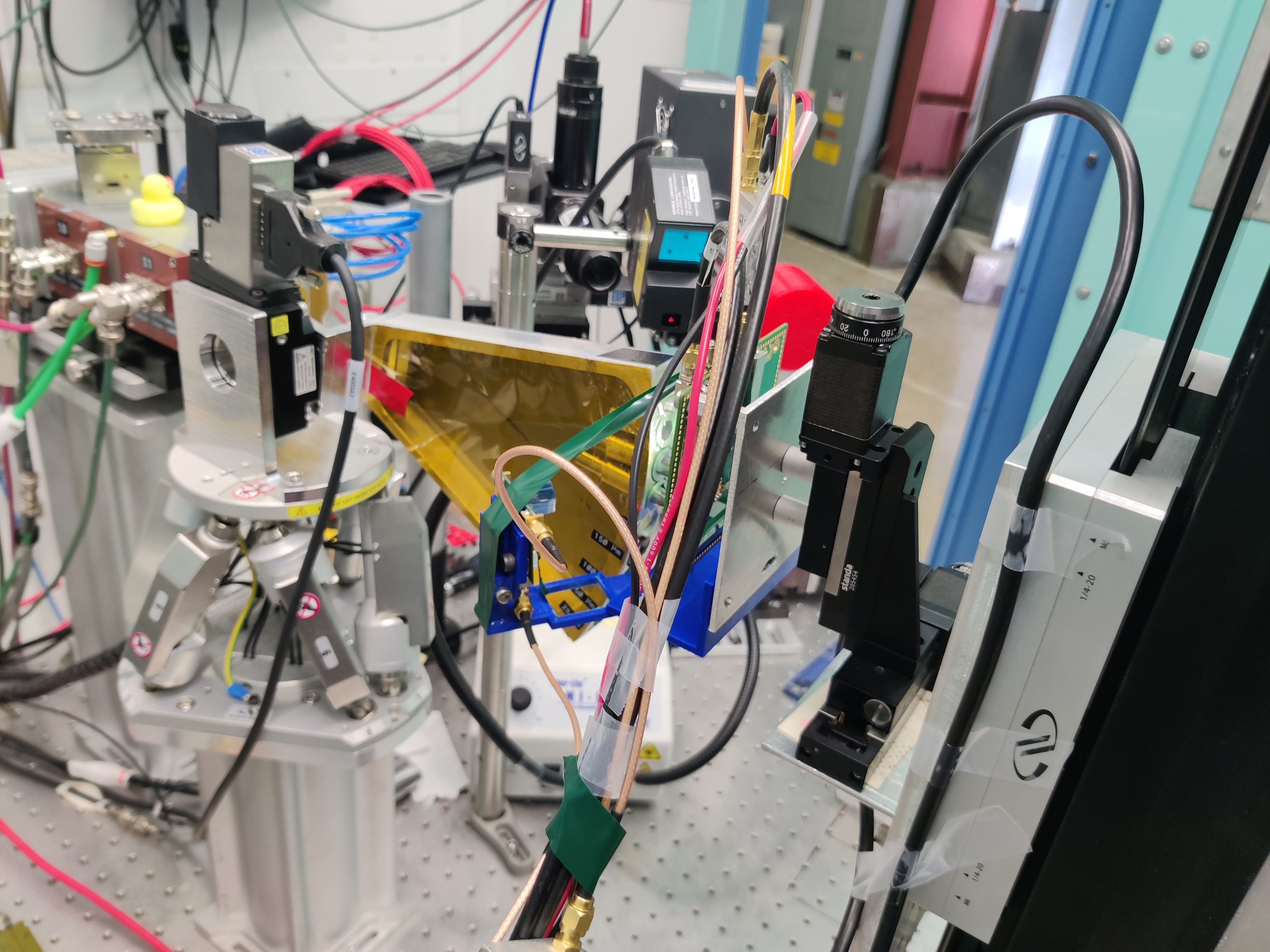}
        \caption{}
    \end{subfigure}
    \caption{(a) 16ch channel analog amplifier readout board used for sensor testing with an FBK AC-LGAD mounted. (b) Data-taking setup at SSRL beam line 7-2. The beam comes from the left side and is collimated by the collimator. The readout board sits on the holder connected to the X-Y micrometer linear stages.}
    \label{fig:setup}
\end{figure}

The tested LGADs are mounted either on a fast single channel analog board (bandwidth $\approx$ \SI{2}{\giga\hertz}) or on a 16 channel fast analog board (bandwidth $\approx$ \SI{1}{\giga\hertz}), shown in Fig.~\ref{fig:setup}(a).
The signal is digitized by a fast oscilloscope or digitizer. 
The experimental setup in the beamline is shown in Fig.~\ref{fig:setup}(b). 
The fast boards have either 1 (schematics in~\cite{seiden2021potential}) or 16 channel(s), and the trans-impedance of the amplifiers is \SI{470}{\ohm} (plus an in-line second-stage amplifier with gain 10) and \SI{5300}{\ohm}, respectively.
The digitizer devices used are: a \SI{2.5}{\giga\hertz} bandwidth \SI{40}{\giga S \per \second}\ oscilloscope\footnote{Keysight Infiniium} (for the single channel boards) 
and a \SI{500}{\mega\hertz}, \SI{5}{\giga S \per \second} 16-channel digitizer\footnote{CAEN DT5742} (for the 16 channel board).
The board sits on a 3D-printed frame mounted on X-Y Thorlabs linear stages for fine adjustment. 
An in-line gas ionization chamber from the SSRL beamline instrumentation measures the beam intensity.
The high voltage (HV) for the sensor bias and low voltage (LV) for powering the board are provided by a tabletop CAEN HV power supply and a low-noise laboratory DC power supply.
The components (HV supply, motors, oscilloscope, and digitizer) are remotely controlled from outside the hutch with a laptop that also controls data collection.

\begin{figure}
    \centering
    \includegraphics[width=0.6\linewidth]{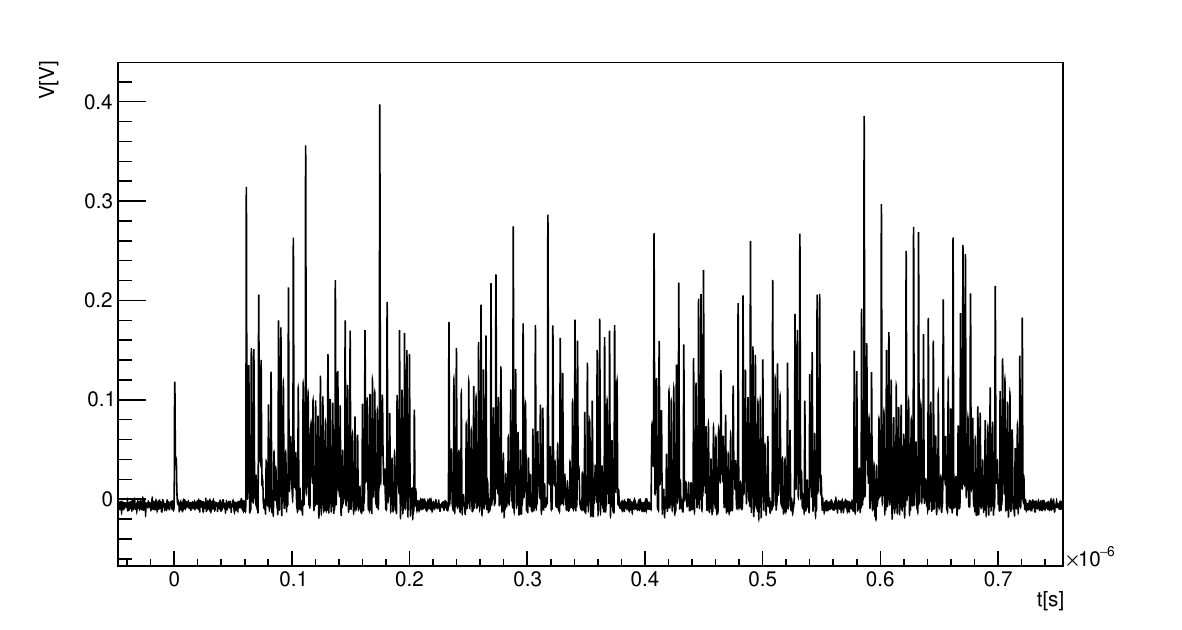}
    \caption{Oscilloscope recorded waveform from an LGAD showing the bunch structure of the SSRL. Four groups of 70 bunches in the beam are separated by an empty interval of a few tens of \SI{}{\nano\second}. The bunch structure is periodic and in phase with the \SI{1.28}{\mega\hertz} synchronizing signal used later in this work to trigger the acquisition.}
    \label{fig:bunch-structure}
\end{figure}

SSRL provides a \SI{1.28}{\mega\hertz} fast synchronization signal in phase with the cyclotron. 
This signal triggered the data acquisition, allowing each event to have the same bunch structure in time.
The beam structure has a window of 70 bunches spaced in time by \SI{2.1}{\nano\second} separated by a few tens of \SI{}{\nano\second} intervals without buckets, which are crucial to evaluate the baseline for each event. 
An example of the recorded signal from an LGAD showing the beam structure is presented in Fig.~\ref{fig:bunch-structure}. At time zero, there's a single separated synchronization bucket that is useful to characterize thicker and slower sensors.
The single pulses are not visible in the figure because the time scale is too large, but pulses with  \SI{2.1}{\nano\second} time distance are fully separated.

\section{Data analysis}
\label{sec:analysis}

The details of the data analysis can be found in~\cite{Mazza:2023col}; a brief summary follows.
For each test condition (sensor type, bias, X-ray beam energy, and position on the sensor), several waveforms were digitized with a length in excess of \SI{100}{\nano\second} triggered by the SSRL synchronizing signal rising edge.
Photons from the \SI{10}{\pico\second} packet in the beam converted inside the LGAD active region will produce a pulse with amplitude proportional to the charges created and multiplied in the device. 
Because there is a significant baseline shift after each event, a baseline correction procedure was devised to remove this effect from the data~\cite{Mazza:2023col}. 
The maximum digitized sample is taken as a simple estimator of the signal amplitude, and the timing was estimated using a CFD method, taking as a reference the packet time separation (\SI{2.1}{\nano\second}).
The quantities are plotted as a histogram and fitted with a Gaussian to extract the central value and resolution.

\section{FBK ``space'' detectors}
\label{sec:FBKspace}

\subsection{Energy response}
\begin{figure}[h]
    \centering
    \begin{subfigure}{0.32\textwidth} 
    \includegraphics[width=\textwidth]
        {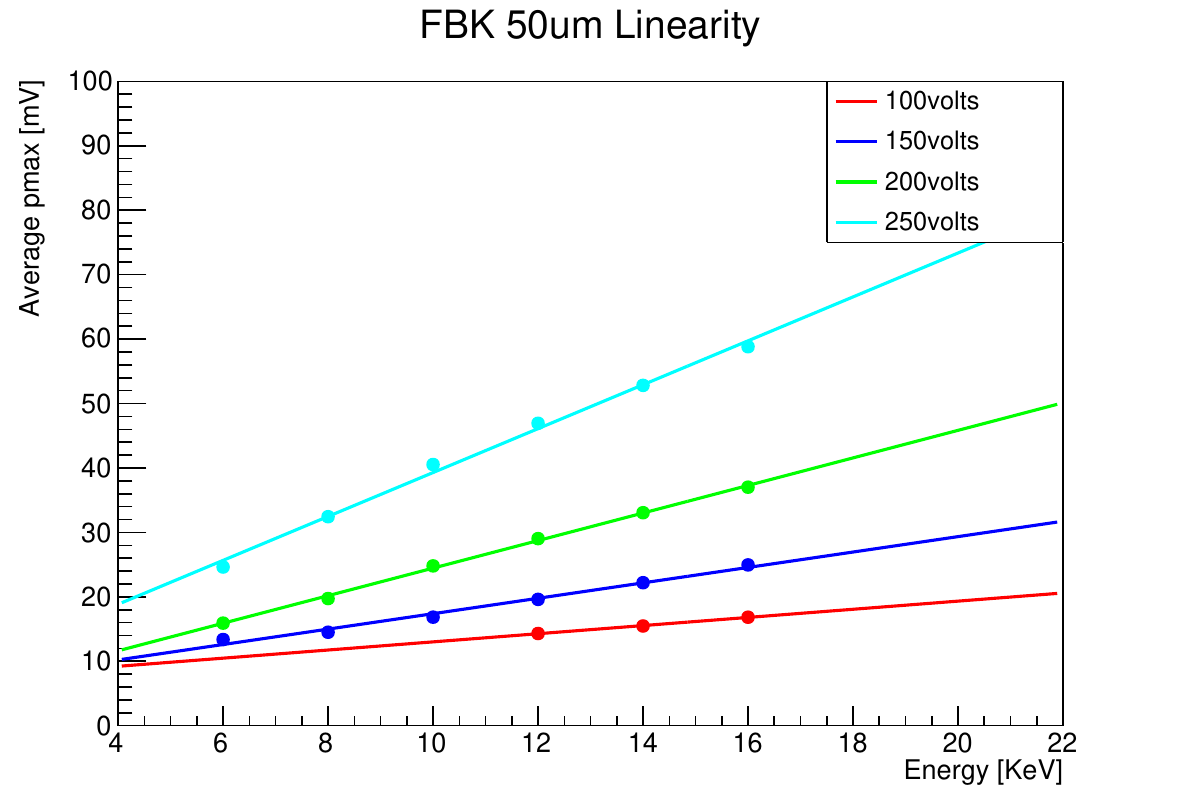}
            \caption{}
    \end{subfigure}
    \begin{subfigure}{0.32\textwidth} 
    \includegraphics[width=\textwidth]
        {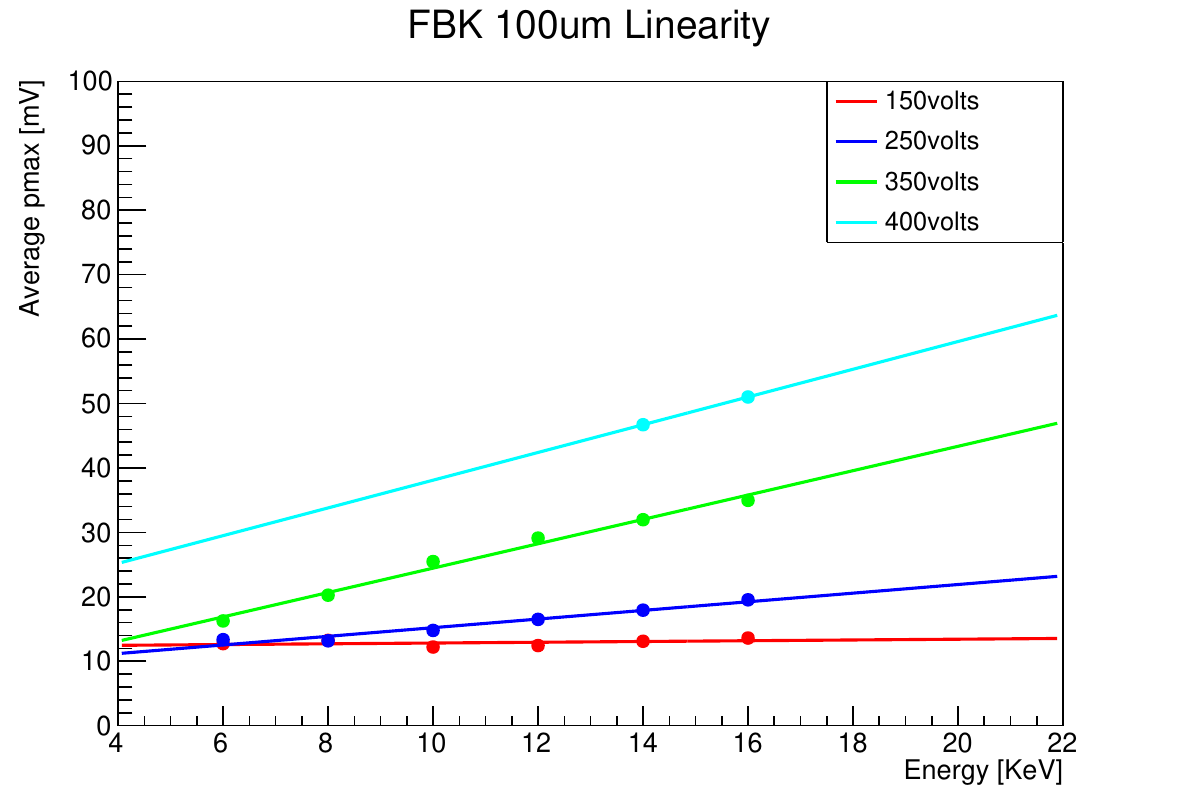}
            \caption{}
    \end{subfigure}
    \begin{subfigure}{0.32\textwidth} 
    \includegraphics[width=\textwidth]
        {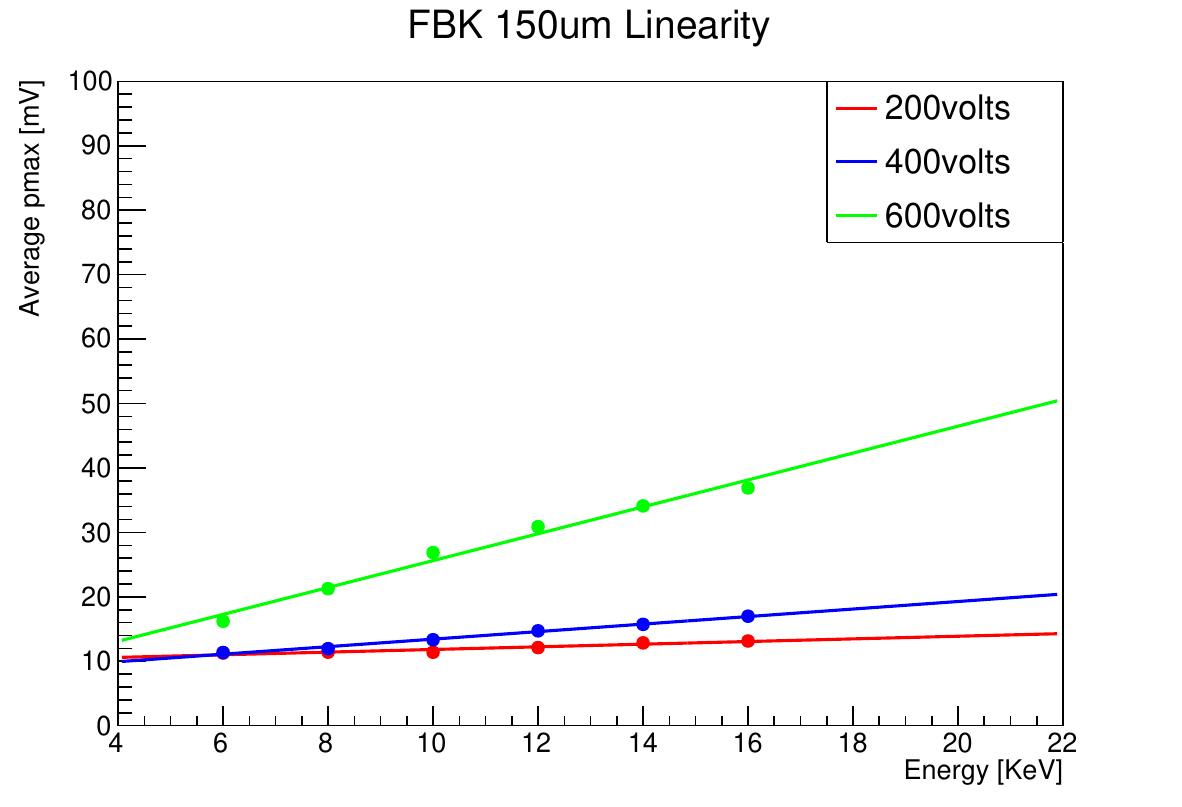}
            \caption{}
    \end{subfigure}
    \begin{subfigure}{0.32\textwidth} 
    \includegraphics[width=\textwidth]
        {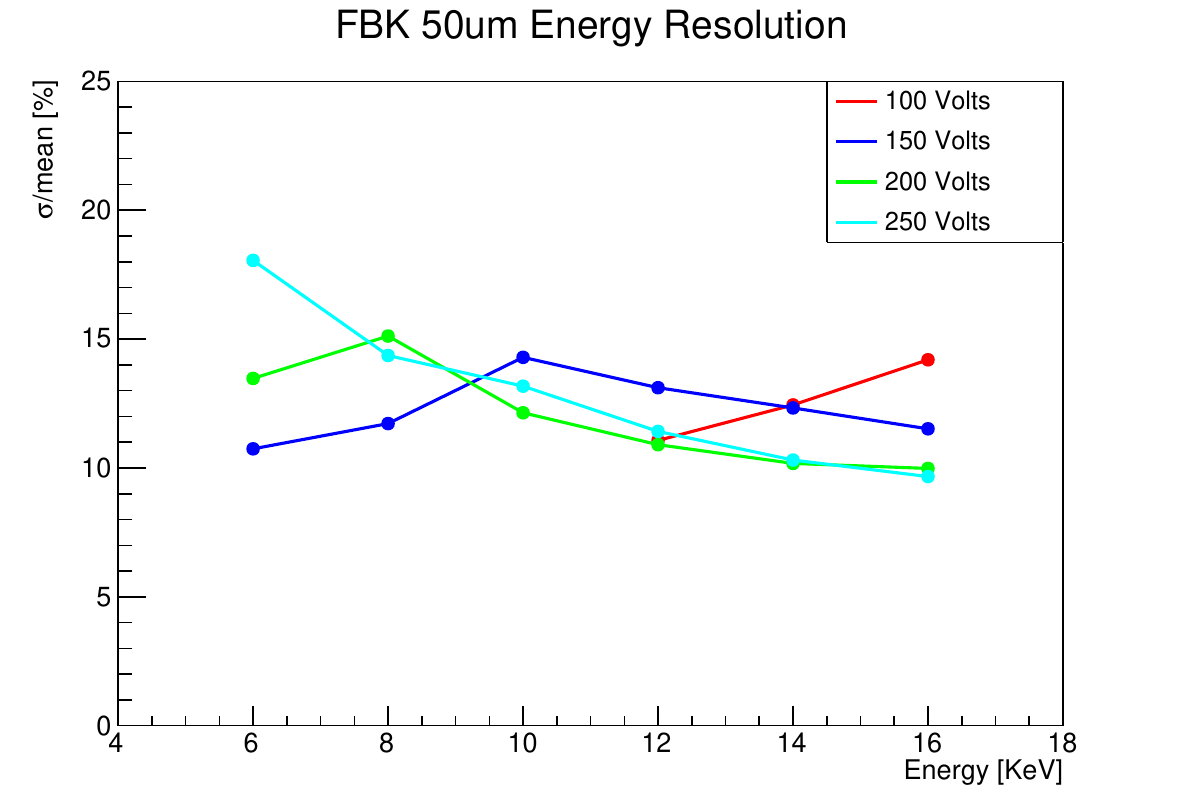}
            \caption{}
    \end{subfigure}
    \begin{subfigure}{0.32\textwidth} 
    \includegraphics[width=\textwidth]
        {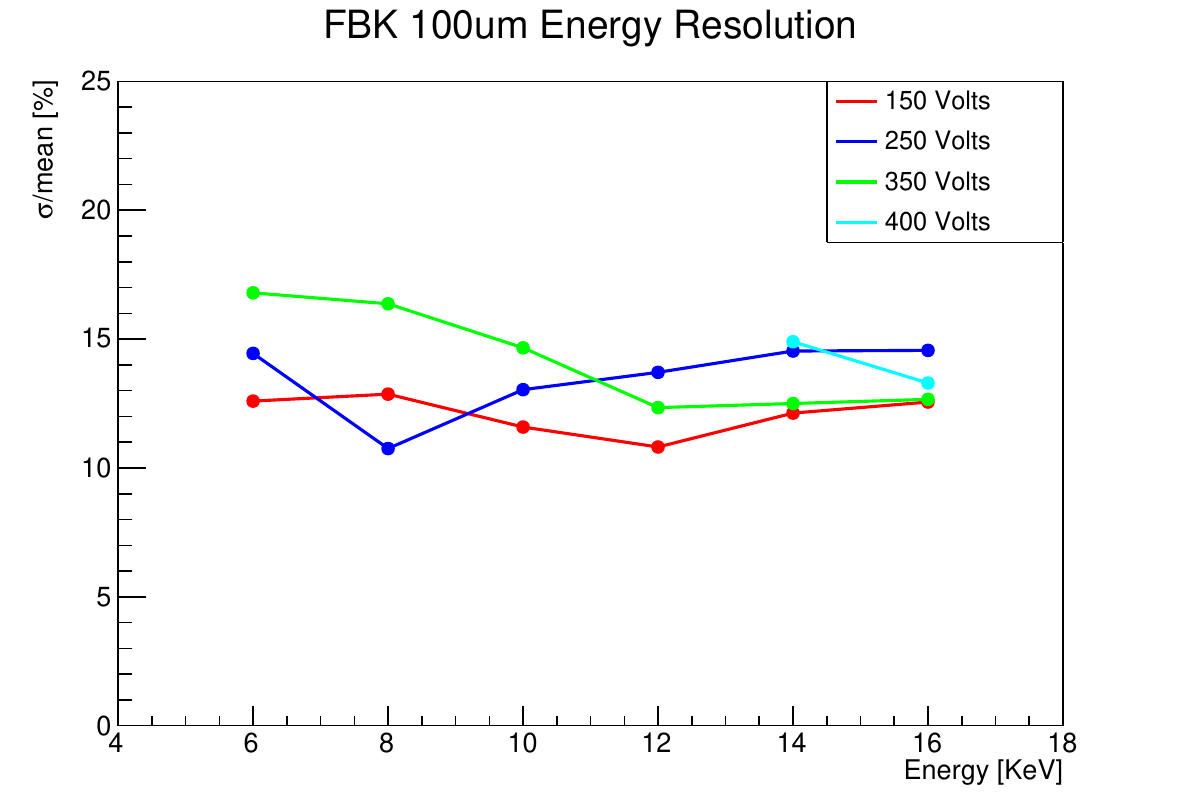}
            \caption{}
    \end{subfigure}
    \begin{subfigure}{0.32\textwidth} 
    \includegraphics[width=\textwidth]
        {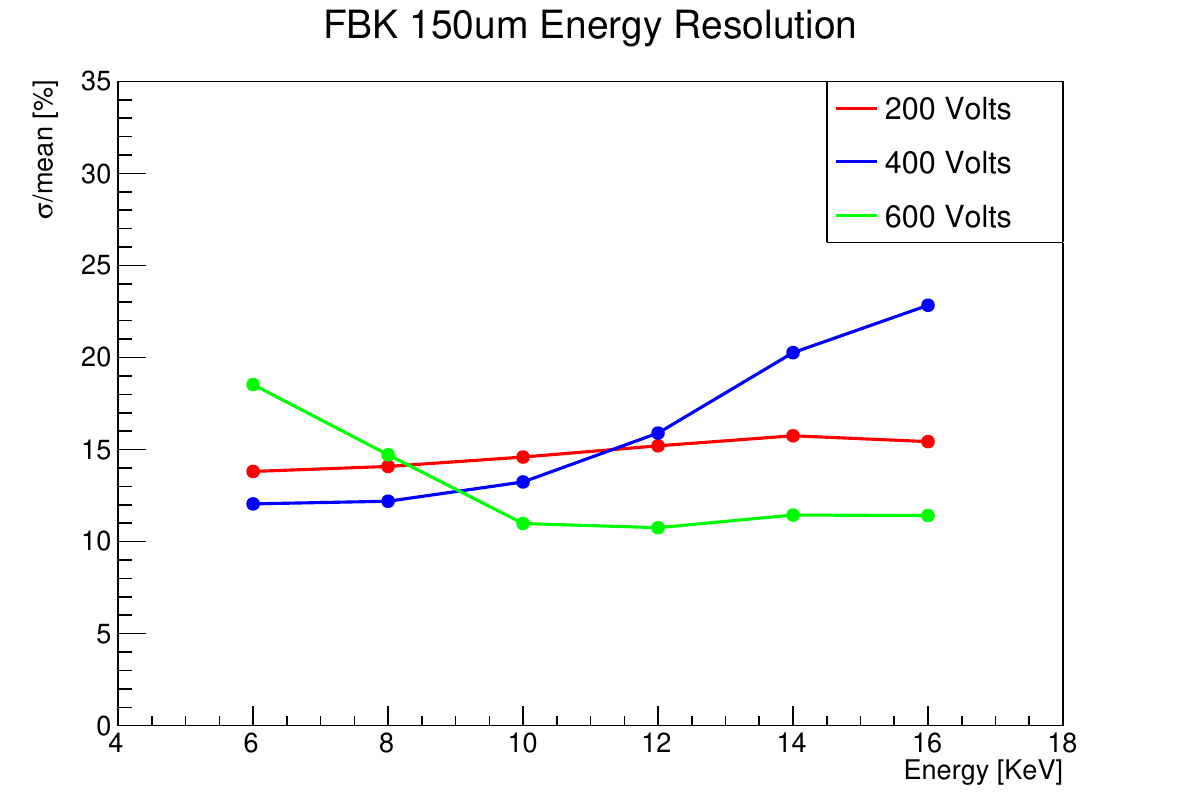}
            \caption{}
    \end{subfigure}
\caption{FBK 'space' sensors energy response at different applied bias voltages for (a) \SI{50}{\micro\meter}, (b) \SI{100}{\micro\meter} and (c) \SI{150}{\micro\meter}. Energy resolution for (c) \SI{50}{\micro\meter}, (d) \SI{100}{\micro\meter} and (e) \SI{150}{\micro\meter}.}
\label{fig:FBK_energy}
\end{figure}

The energy response is calculated as the mean of a Gaussian fit on the pulse amplitude distribution after baseline correction. Similarly, the energy resolution is the standard deviation of the fitted Gaussian. The measurement is done for all X-ray energies and applied bias voltages. The energy linearity and resolution are shown in Fig.~\ref{fig:FBK_energy}, top and bottom, respectively.
As the plots show, the sensors' response is linear as expected, with an amplitude between 10~mV and 50~mV for all tested conditions. Given that the noise for the readout board employed is around 2~mV, a signal-to-noise ratio of ten is achieved for most of the running conditions.
The energy resolution is between 10\% and 15\% for most running conditions, a result slightly worse but similar to what was measured for another type of detector in the previous measurement campaign~\cite{Mazza:2023col}.

\subsection{Time response}
\begin{figure}[h]
    \centering
    \begin{subfigure}{0.32\textwidth} 
    \includegraphics[width=\textwidth]
        {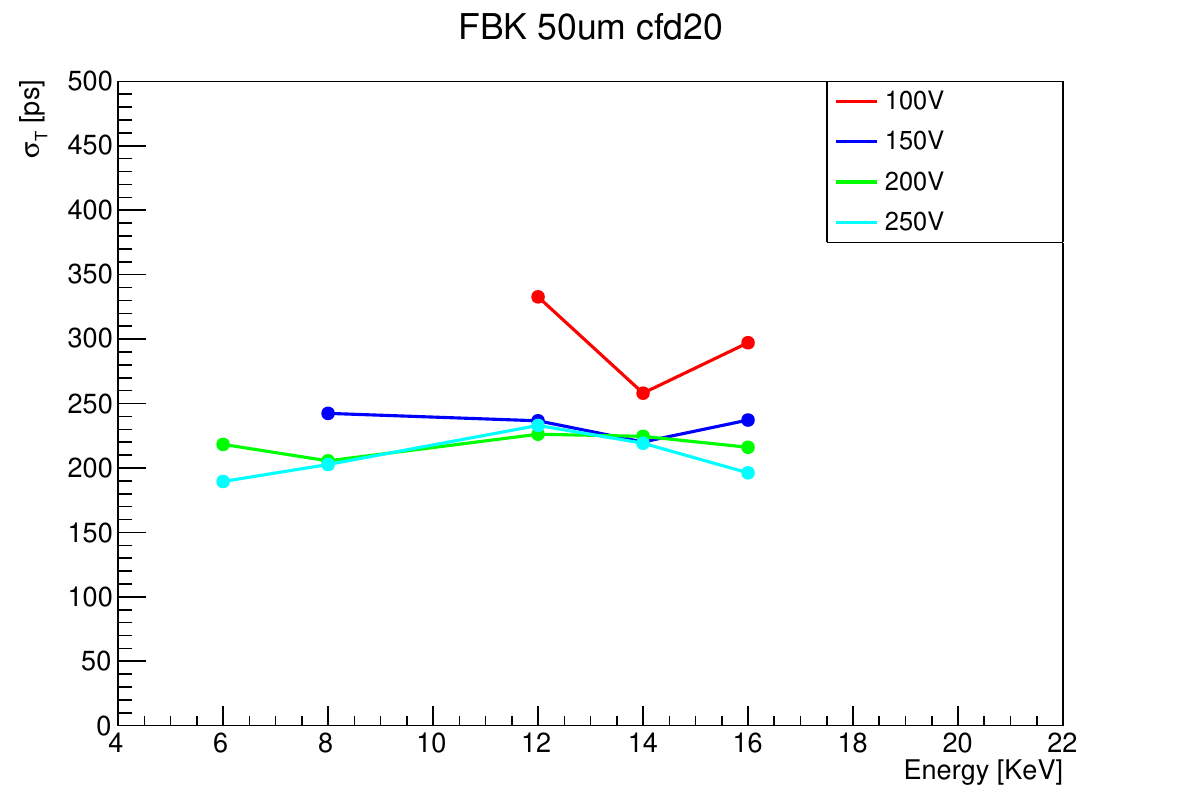}
            \caption{}
    \end{subfigure}
    \begin{subfigure}{0.32\textwidth} 
    \includegraphics[width=\textwidth]
        {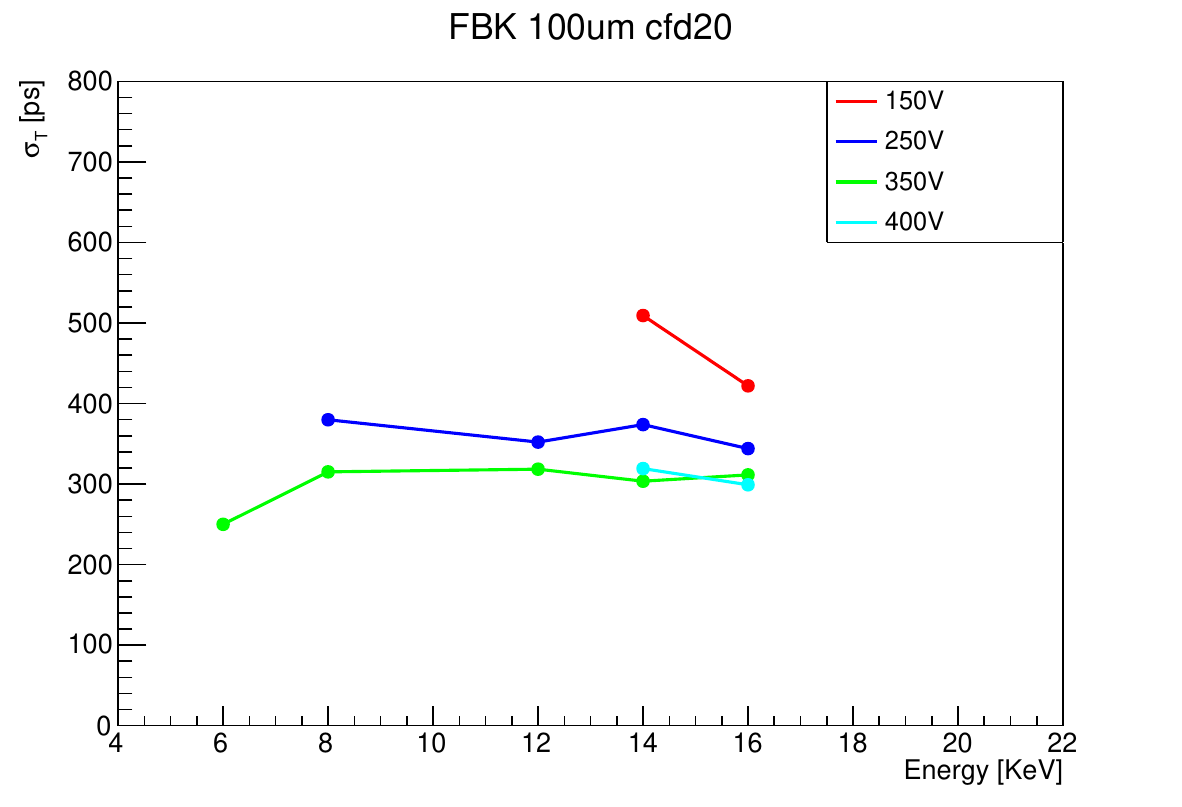}
            \caption{}
    \end{subfigure}
    \begin{subfigure}{0.32\textwidth} 
    \includegraphics[width=\textwidth]
        {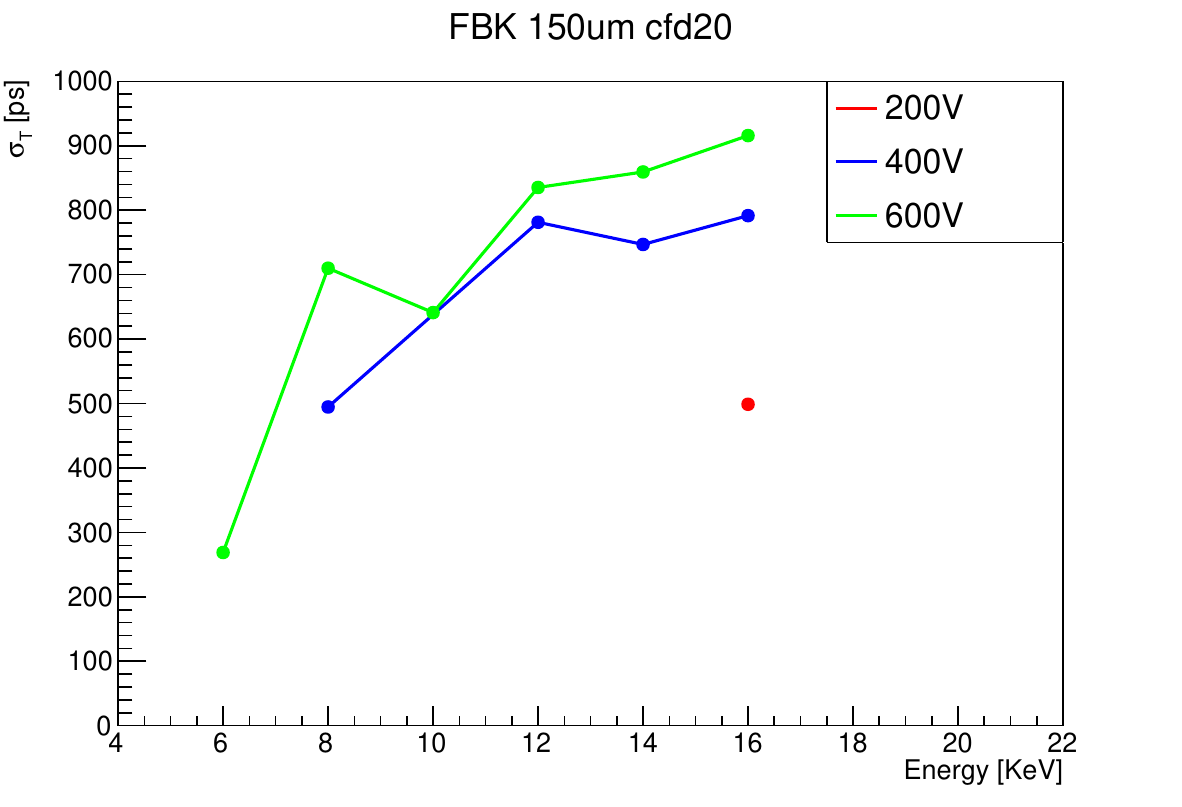}
            \caption{}
    \end{subfigure}
\caption{FBK 'space' sensors time resolution at different applied bias voltages for (a) \SI{50}{\micro\meter}, (b) \SI{100}{\micro\meter} and (c) \SI{150}{\micro\meter}.}
\label{fig:FBK_time}
\end{figure}

The time resolution is calculated as the Gaussian sigma of the time difference between the first and the fifth 
pulses in the bunch train, divided by $\sqrt2$.
The results for the three thicknesses are shown in Fig.~\ref{fig:FBK_time}.
The time resolution of the \SI{50}{\micro\meter} detector is around 200~ps, which is worse than the results (around 100~ps) from the previous test beam using an HPK device. This can be explained by the difference in sensor design and marginally higher thickness (\SI{55}{\micro\meter} instead of $<$\SI{50}{\micro\meter}).
The results for the thicker sensor are increasingly, as expected, worse due to the increased drift time of the deposited charge.

\subsection{Waveform analysis}
\begin{figure}[h]
    \centering
    \begin{subfigure}{0.32\textwidth} 
    \includegraphics[width=\textwidth]
        {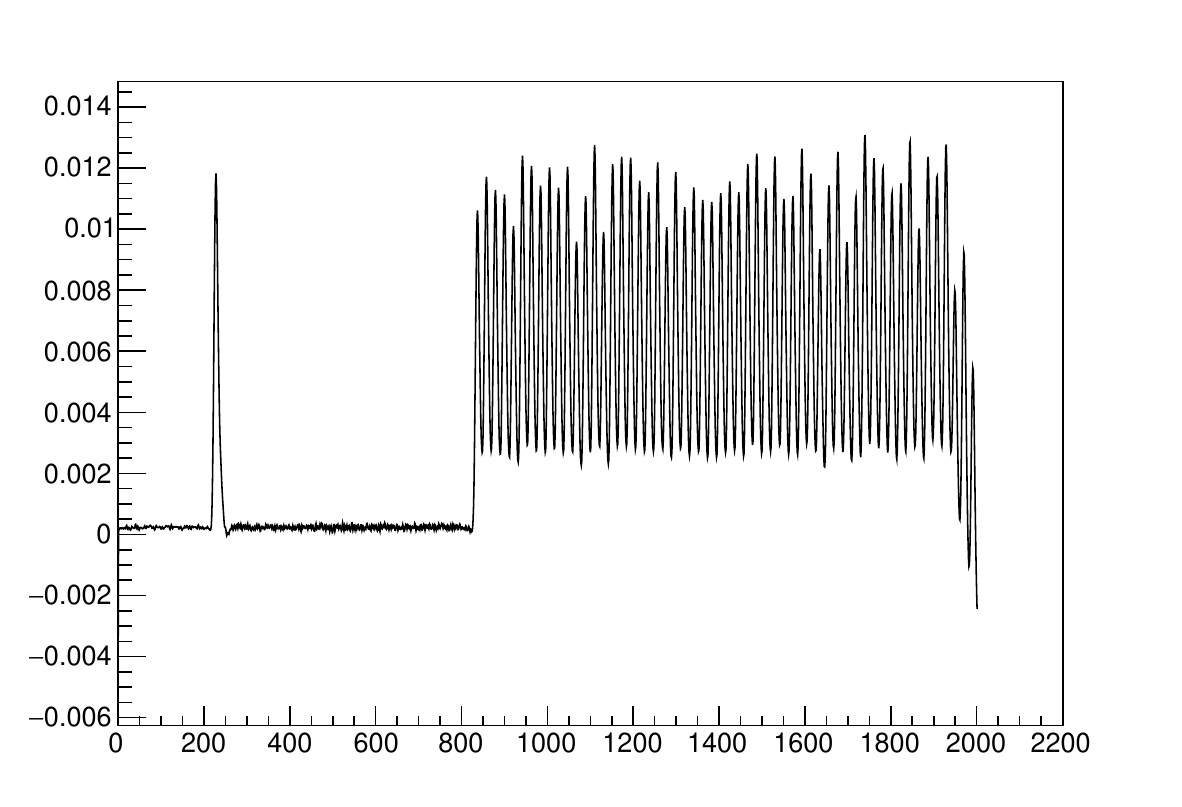}
            \caption{}
    \end{subfigure}
    \begin{subfigure}{0.32\textwidth} 
    \includegraphics[width=\textwidth]
        {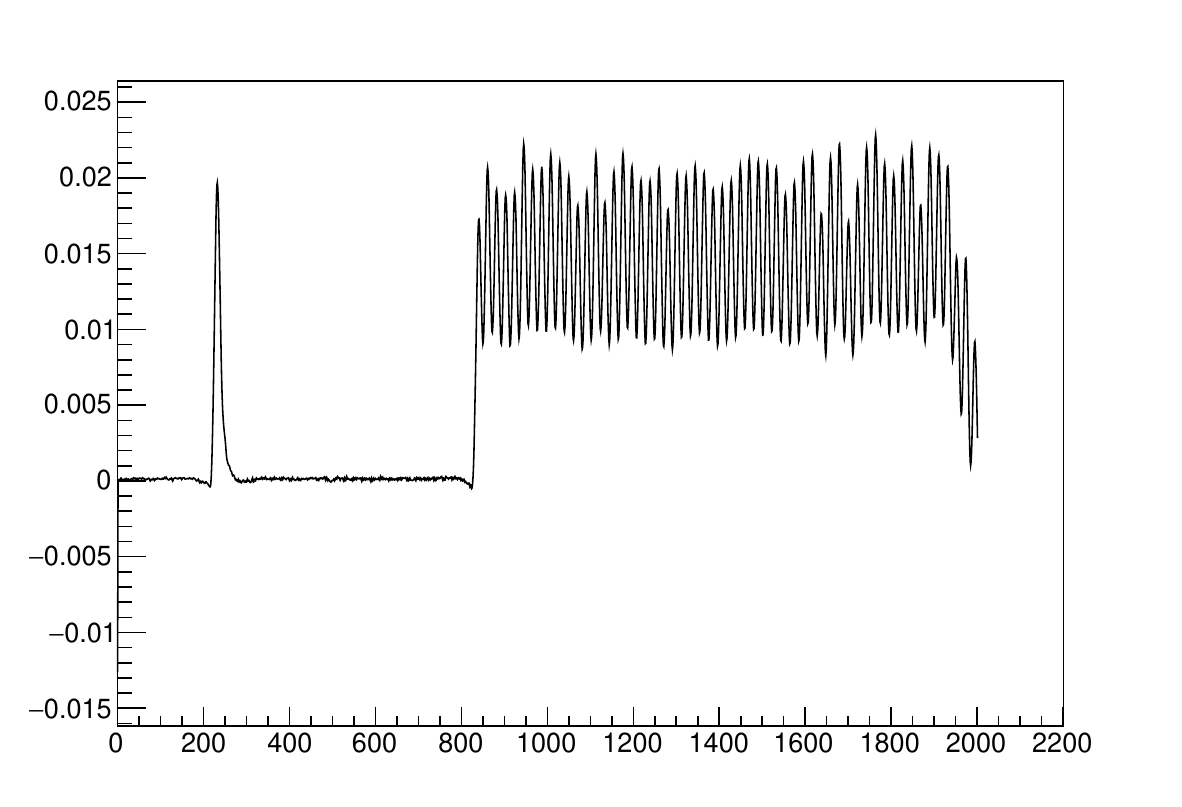}
            \caption{}
    \end{subfigure}
    \begin{subfigure}{0.32\textwidth} 
    \includegraphics[width=\textwidth]
        {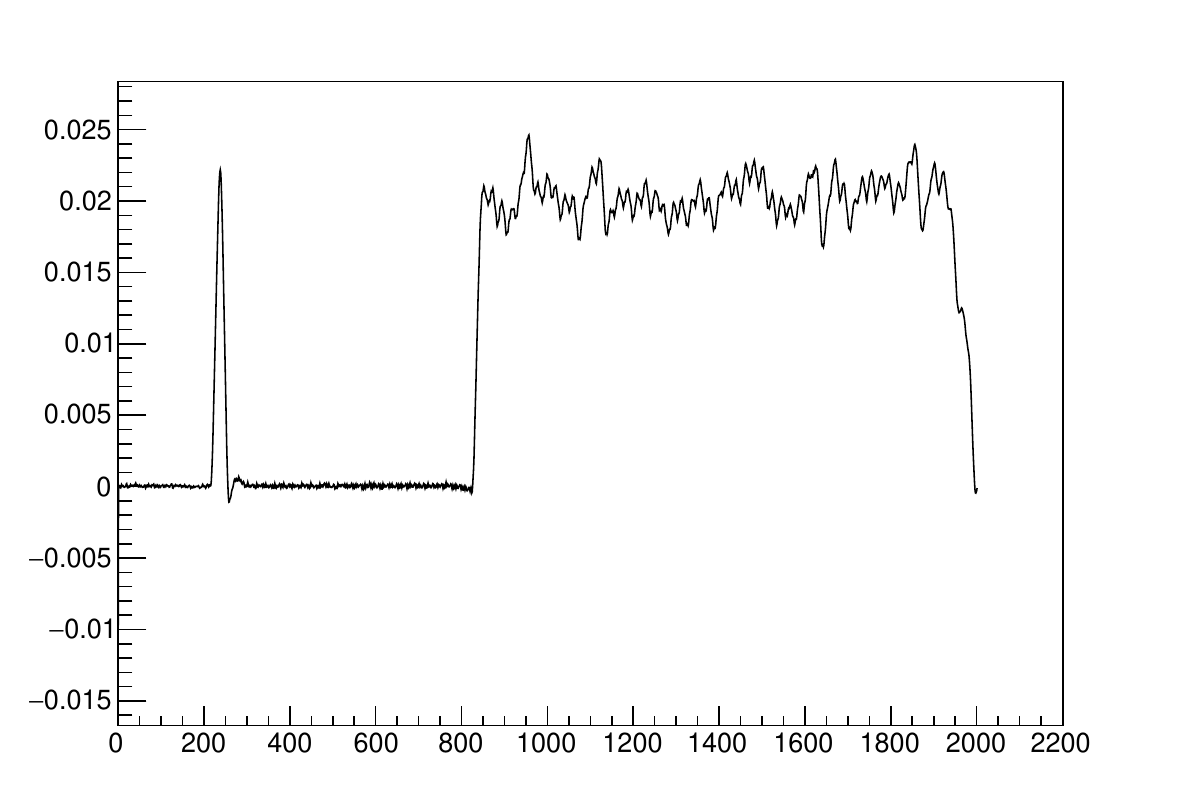}
            \caption{}
    \end{subfigure}
\caption{FBK 'space' sensors average waveform of a bunch train section for (a) \SI{50}{\micro\meter}, (b) \SI{100}{\micro\meter} and (c) \SI{150}{\micro\meter}.}
\label{fig:FBK_train}
\end{figure}

\begin{figure}[h]
    \centering
    \begin{subfigure}{0.32\textwidth} 
    \includegraphics[width=\textwidth]
        {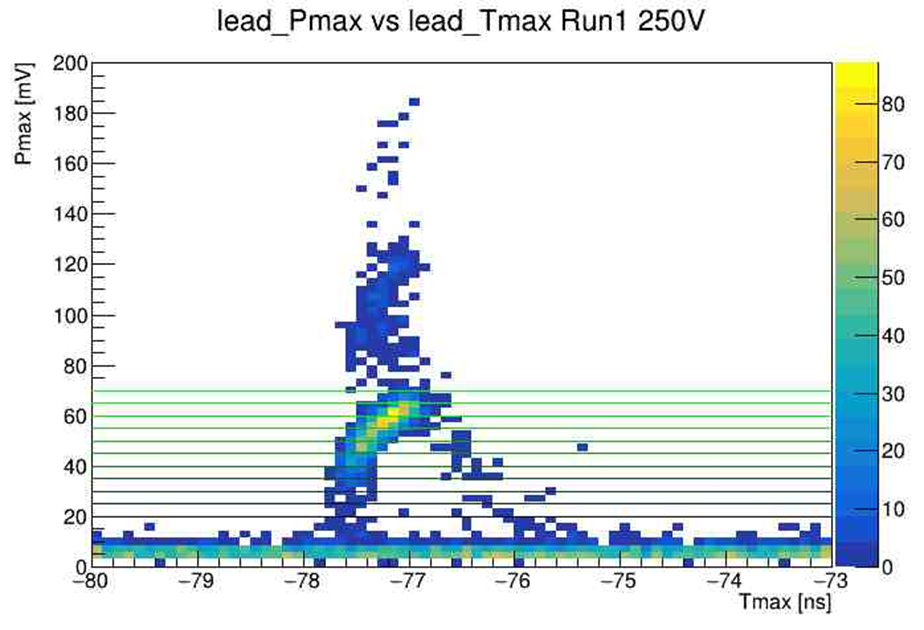}
            \caption{}
    \end{subfigure}
    \begin{subfigure}{0.32\textwidth} 
    \includegraphics[width=\textwidth]
        {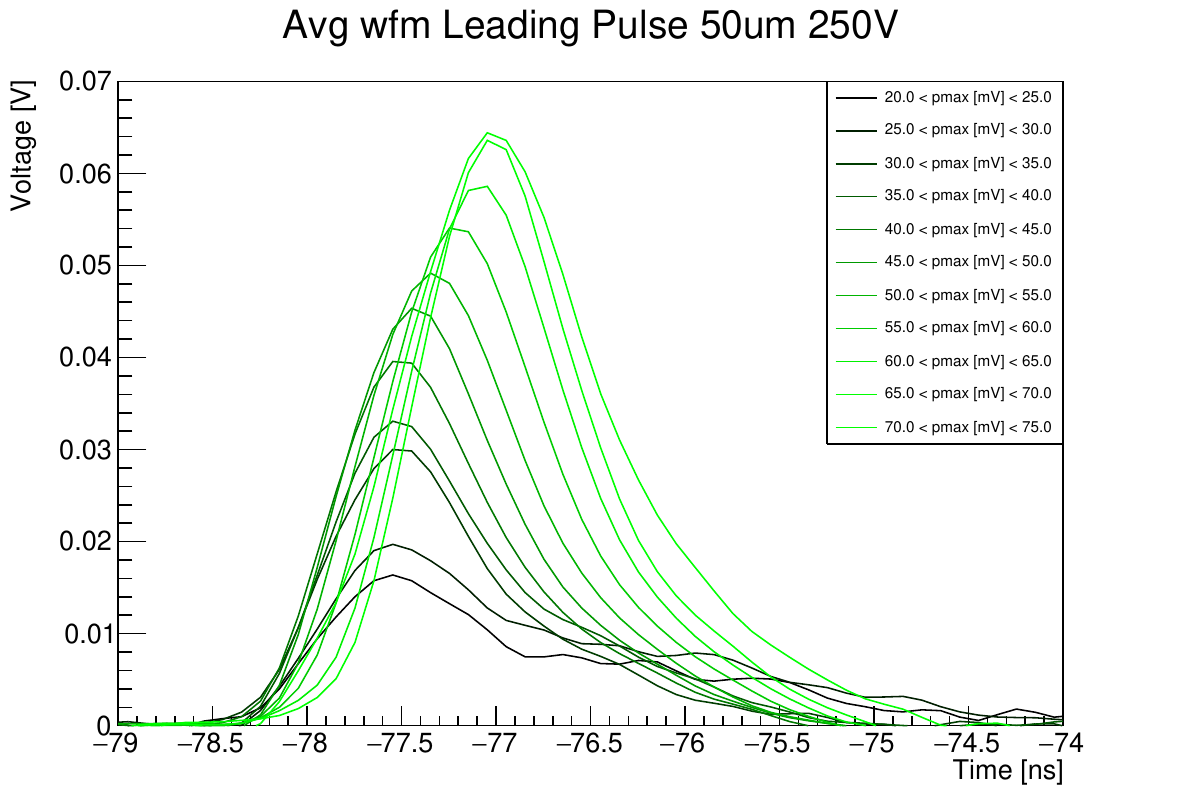}
            \caption{}
    \end{subfigure}
    \begin{subfigure}{0.32\textwidth} 
    \includegraphics[width=\textwidth]
        {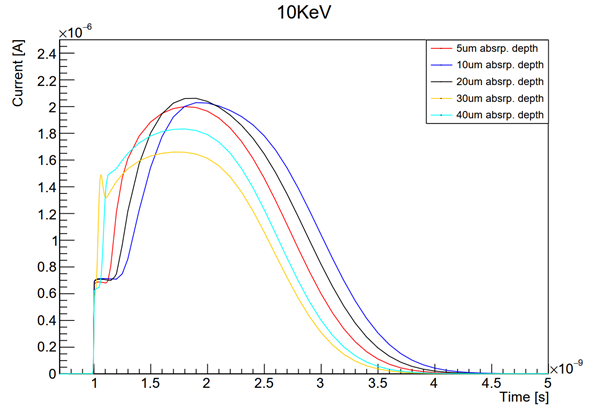}
            \caption{}
    \end{subfigure}
\caption{\SI{50}{\micro\meter} FBK 'space' sensors (a) Pmax vs Tmax distribution, (b) average waveforms in the Pmax intervals shown in (a) and (c) TCAD simulation of charge collection as a function of charge deposition depth from the top contact surface.}
\label{fig:FBK_pulse50}
\end{figure}

The average bunch train waveform for the three devices can be seen in Fig.~\ref{fig:FBK_train}, the \SI{50}{\micro\meter} sensor in (a) can resolve the bunch structure. However, the pulses are significantly overlapping for the \SI{100}{\micro\meter} (b) sensor and the \SI{150}{\micro\meter} sensor (c).
A detailed study was done only on the single separated pulses before the bunch train to avoid waveform overlap. 
The results for the \SI{50}{\micro\meter} sensor is shown in Fig.~\ref{fig:FBK_pulse50}, the plots show in (a) the Pulse maximum (Pmax) vs Time of the pulse maximum (Tmax), in (b) the average waveform for selected bands of Pmax and in (c) a TCAD sentaurus waveform simulation of a same-thickness LGAD for different X-ray conversion depth.
\FloatBarrier
In the distribution of Fig.~\ref{fig:FBK_pulse50} (a), a few different populations of events are divided as follows.
Below \SI{20}{\milli\volt} of $\pmax$ is noise or events likely coming from Compton interaction between X-ray and Silicon bulk. 
Between \SI{20}{\milli\volt} and \SI{70}{\milli\volt} of $\pmax$ is the main group of events for \SI{16}{\kilo\electronvolt} X-rays.
Fig.~\ref{fig:FBK_pulse50} (b) shows the average color-coded waveforms for event selections in $\pmax$ corresponding to the lines drawn in Fig.~\ref{fig:FBK_pulse50} (a).
It can be seen that pulses with a larger $\pmax$ have a delayed initial rising edge.

The depth of X-ray absorption causes the variation in $\tmax$, as expected from the simulation in Fig.~\ref{fig:FBK_pulse50} (c).
The events for $\pmax$ over \SI{150}{\milli\volt} are for double photon absorption. 
Furthermore, a slight gain loss is expected when the photon interacts close to the gain layer, as seen in Fig.~\ref{fig:FBK_pulse50} (c).
This can be explained as the effect of the charge density on the electric field, which is reduced, and therefore, the gain is reduced. This mechanism is called gain suppression~\cite{Braun:2024sbi}.
When the charge is deposited deeper into the detector, the electron lateral drift reduces the charge density in the gain layer.
However, the observed direct correlation between $\tmax$ and $\pmax$ in the main distribution hints at a larger gain reduction than the one predicted in the simulation.

\begin{figure}[h]
    \centering
    \begin{subfigure}{0.32\textwidth} 
    \includegraphics[width=\textwidth]
        {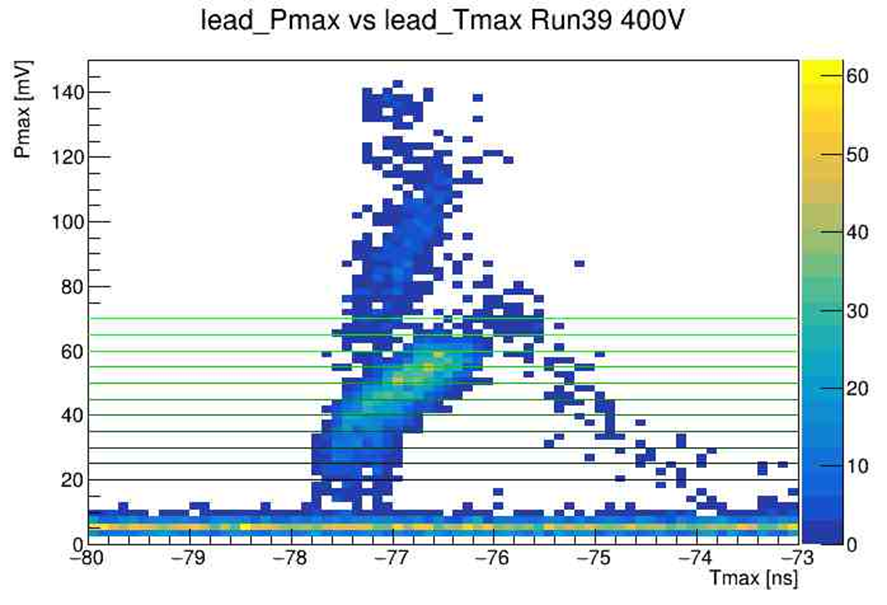}
            \caption{}
    \end{subfigure}
    \begin{subfigure}{0.32\textwidth} 
    \includegraphics[width=\textwidth]
        {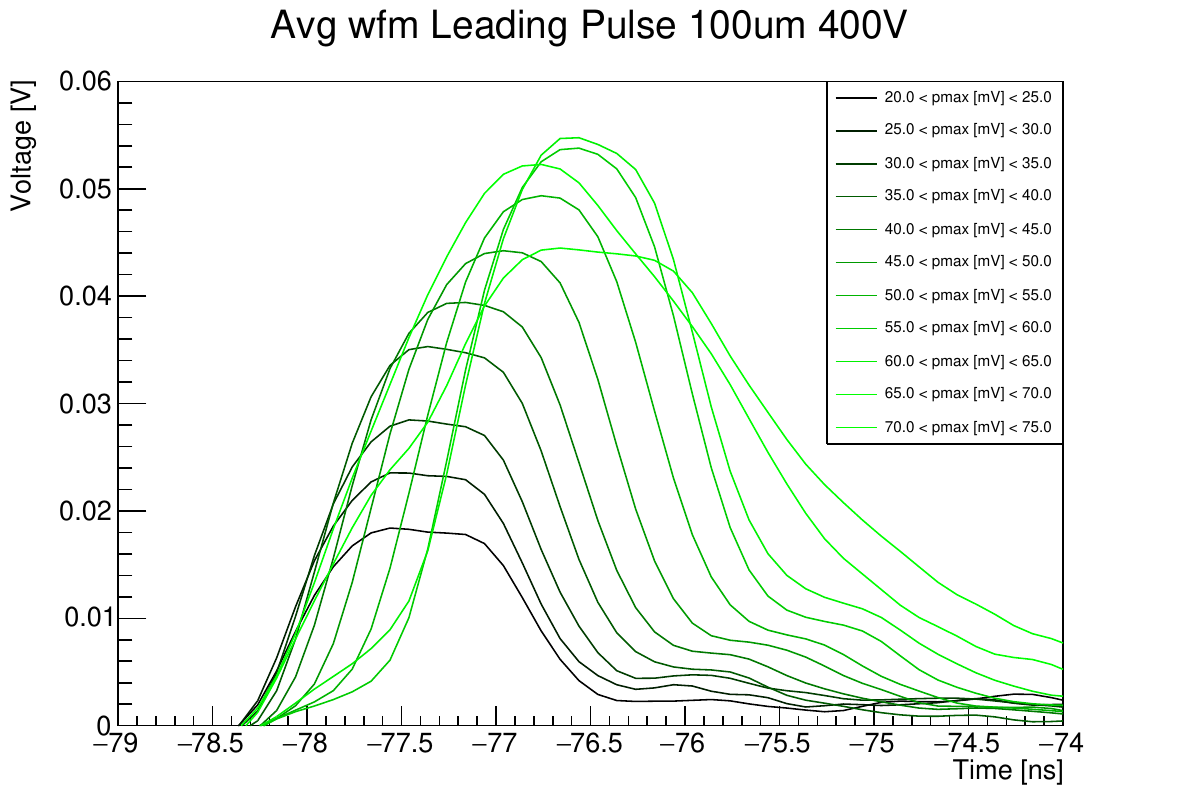}
            \caption{}
    \end{subfigure}
    \begin{subfigure}{0.26\textwidth} 
    \includegraphics[width=\textwidth]
        {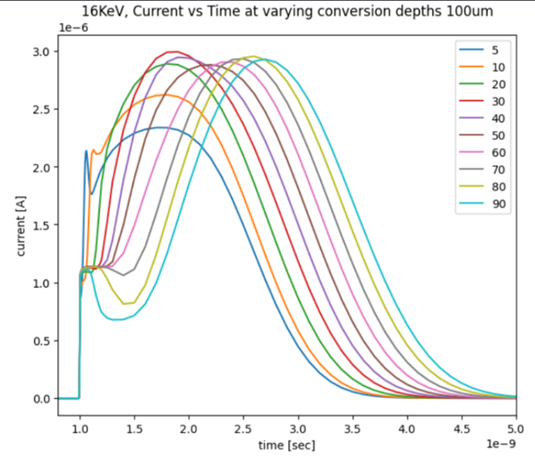}
            \caption{}
    \end{subfigure}
\caption{\SI{100}{\micro\meter} FBK 'space' sensors (a) Pmax vs Tmax distribution, (b) average waveforms in the Pmax intervals shown in (a) and (c) TCAD simulation of charge collection as a function of charge deposition depth from the top contact surface.}
\label{fig:FBK_pulse100}
\end{figure}

\begin{figure}[h]
    \centering
    \begin{subfigure}{0.32\textwidth} 
    \includegraphics[width=\textwidth]
        {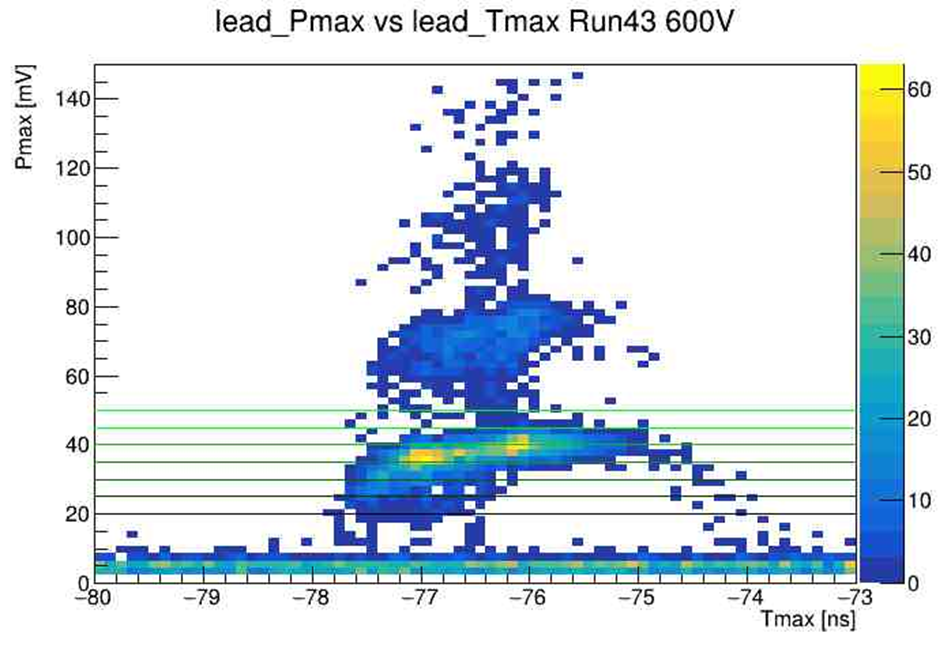}
            \caption{}
    \end{subfigure}
    \begin{subfigure}{0.32\textwidth} 
    \includegraphics[width=\textwidth]
        {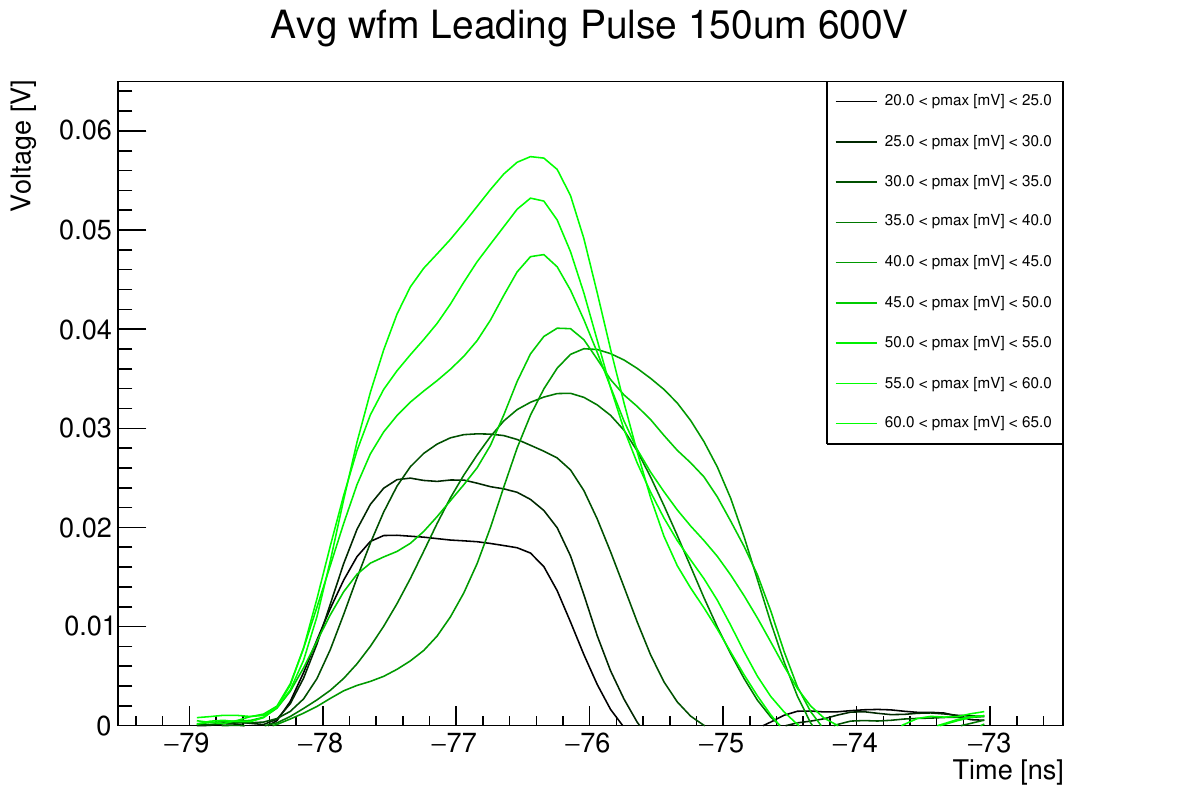}
            \caption{}
    \end{subfigure}
    \begin{subfigure}{0.26\textwidth} 
    \includegraphics[width=\textwidth]
        {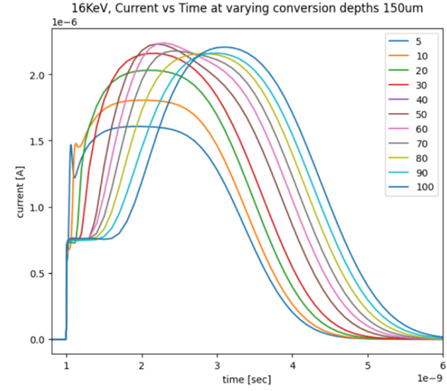}
            \caption{}
    \end{subfigure}
\caption{\SI{150}{\micro\meter} FBK 'space' sensors (a) Pmax vs Tmax distribution, (b) average waveforms in the Pmax intervals shown in (a) and (c) TCAD simulation of charge collection as a function of charge deposition depth from the top contact surface.}
\label{fig:FBK_pulse150}
\end{figure}

The same study is repeated for the \SI{100}{\micro\meter} (Fig.~\ref{fig:FBK_pulse100}) and the \SI{150}{\micro\meter} (Fig.~\ref{fig:FBK_pulse150}). 
The behavior is similar to the \SI{50}{\micro\meter} device, however, with increased delay (as expected) for the onset of gain for the backside conversion.
The difference in gain suppression is also notable; it does not indefinitely depend on the conversion depth but saturates when the X-ray converts at about \SI{50}{\micro\meter} inside the detector.
This behavior, although not perfectly agreeing with the simulation, is also seen in the average pulses from the SSRL data.

\section{AC-LGADs and TI-LGADs}
\label{sec:acti}
\begin{figure}[h]
    \centering
    \begin{subfigure}{0.47\textwidth} 
    \includegraphics[width=\textwidth]
        {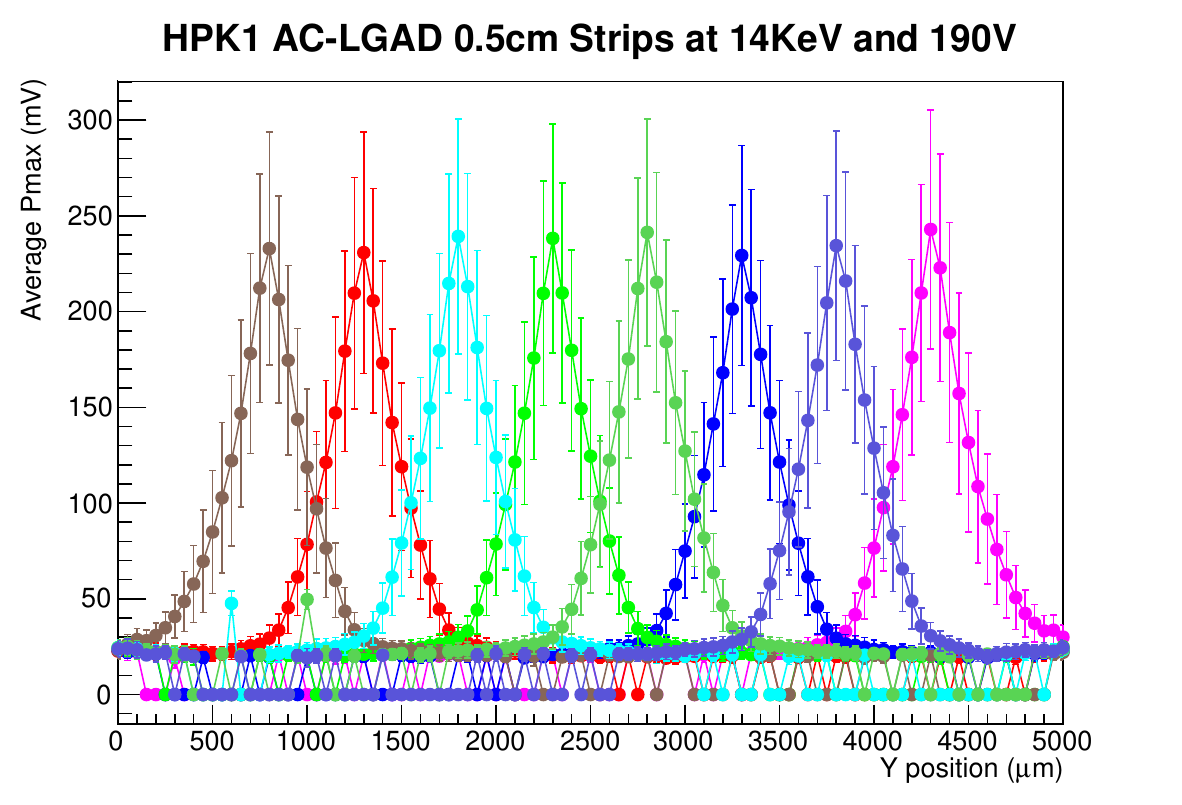}
        \caption{}
    \end{subfigure}
    \begin{subfigure}{0.47\textwidth} 
    \includegraphics[width=\textwidth]
        {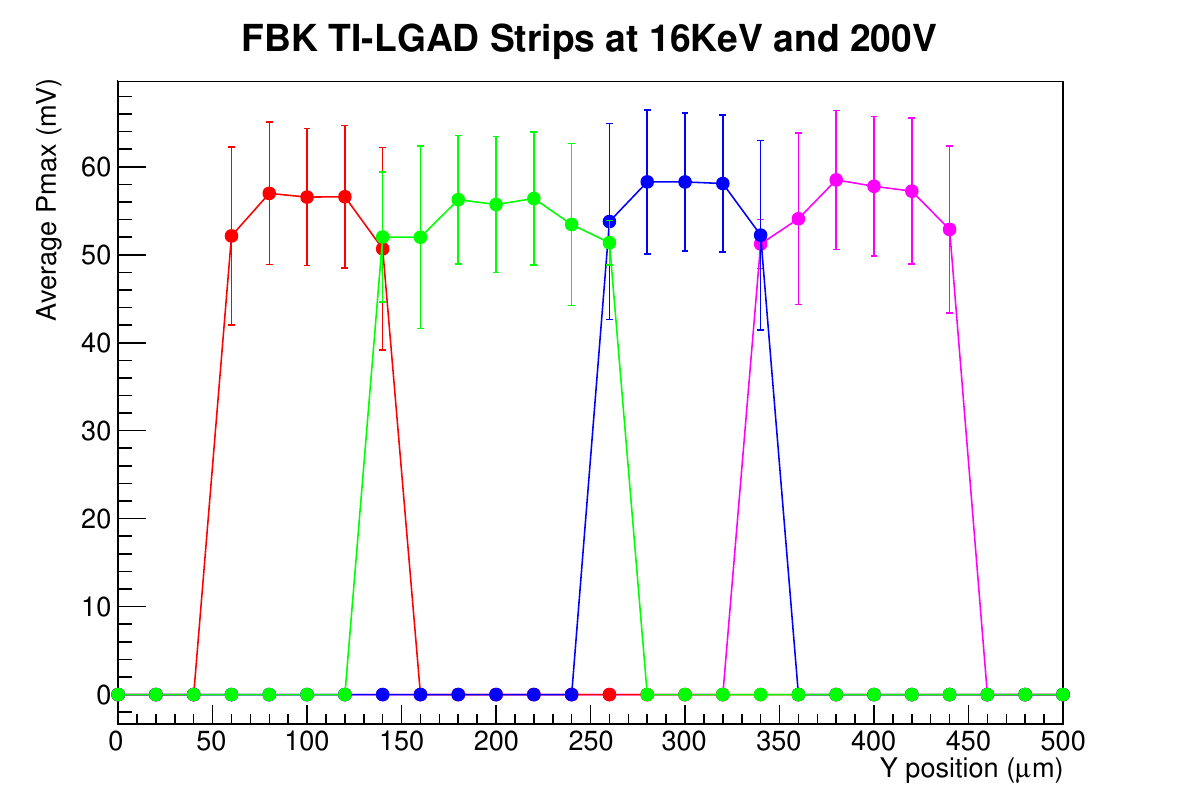}
        \caption{}
    \end{subfigure}
\caption{Pmax response of HPK AC strips as a function of position for HPK1 E240 strip board 0.5cm sensor strip board 0.5cm sensor. The different colors indicate different strips. (b) FBK TI-LGADs beam scans of the strip detector.}
\label{fig:HPK_strips}
\end{figure}

\begin{figure}[h]
    \centering
    \begin{subfigure}{0.5\textwidth} 
    \includegraphics[width=\textwidth]
        {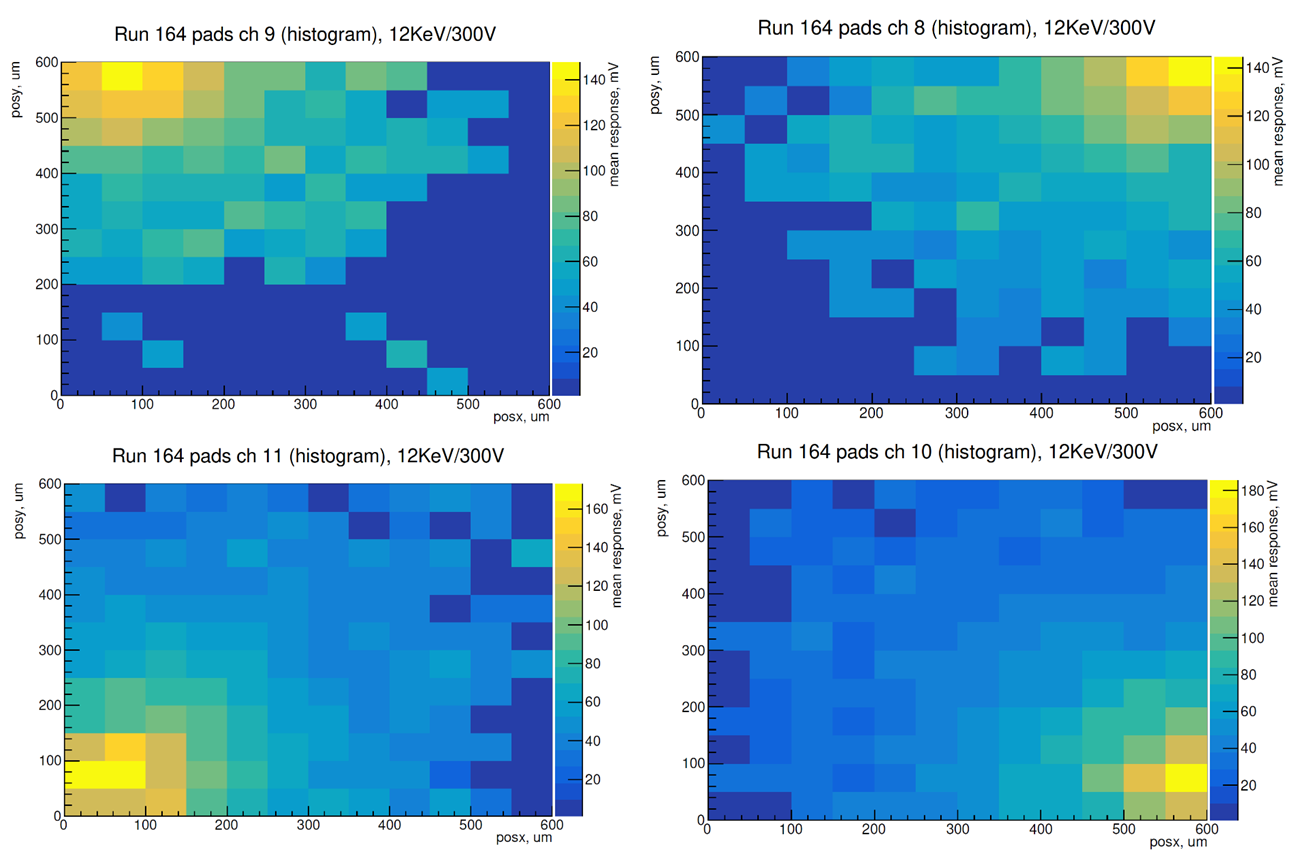}
            \caption{}
    \end{subfigure}
\caption{2D beam scan of FBK AC-LGAD sensor with metal square pads electrodes and pitch of \SI{500}{\micro\meter}. Each plot represents the response of one of the four electrodes at the corner of the image.}
\label{fig:FBK_AC}
\end{figure}

AC-LGADs were also tested, using the narrow focus beam of 7-2.
The HPK strip sensor (Fig.~\ref{fig:LGADS}, (b)) was scanned perpendicular to the strips across the entire sensor, roughly in the sensor's vertical center. 
The data was collected using the FNAL 16ch board and the CAEN 16ch digitizer.
The results are shown in Fig.~\ref{fig:HPK_strips} (a). 
To measure the point in the plots, for each strip and position, a distribution of the Pmax is formed and fitted with a Gaussian; the mean and sigma are the points and their uncertainty. 
The different colors in the plots indicate different strips on the sensor.
The characteristic charge-sharing behavior of AC-LGADs is evident in the plot; the signal is maximum near the strip position and decreases with distance. AC-LGADs can achieve a position resolution significantly lower than the pitch by combining the information of several electrodes in the reconstruction~\cite{MENZIO2024169526}.
The distributions are in agreement with those obtained in the laboratory with a focused TCT laser~\cite{Bishop:2024sjy}.
Fig.~\ref{fig:FBK_AC} contains the response as a function of the position of all four square metal pads of Fig.~\ref{fig:LGADS} (c). The area between the four electrodes was scanned, and each 2D distribution corresponds to a different metal pad.
The charge-sharing behavior is on display for this sensor as well.

A similar study was made on FBK TI-LGAD strips, Fig.~\ref{fig:LGADS} (e).
The results are shown in Fig.~\ref{fig:HPK_strips} (d). No cross-talk is observed between electrodes as expected for a DC-connected LGAD. 
The inter-pad gap for this kind of device is 5-10~$\mu$m or even less than 5~$\mu$m, depending on the production parameters~\cite{9081916,s23136225}; therefore, no gap is observed as the beam width is around 30~$\mu$m and cannot resolve such a small feature.
The response is homogeneous across the sensor as expected.

\section{Conclusions}
\label{sec:conclusions}

X-ray detection of silicon LGAD devices of several types was studied at the Stanford Synchrotron Radiation  LightSource (SSRL). 
Single-pad devices of 50~$\mu$m, 100~$\mu$m, and 150~$\mu$m from FBK were tested and show an energy resolution between 10\% and 15\% in the X-ray energy range.
The tested 50~$\mu$m-thin devices easily resolved in time the 500~MHz repetition rate of the beam line, while the thicker devices cannot.
The time resolution was estimated to be between 100~ps and 1000~ps after time-walk correction, depending on the thickness; this value is significantly worse than in the case of minimum ionizing particles.
The effect of X-ray conversion depth was also studied using average waveforms selected in regions of Pmax, the results were verified with TCAD simulations.
AC-LGADs and TI-LGADs from HPK and FBK were also tested using the focused X-ray beam, and a behavior similar to MIP injection and TCT laser scanning was observed.

{
\small
\acknowledgments
This work was supported by the United States Department of Energy grants DE-SC0010107, DE-FG02-04ER41286, and DE‐FG02‐97ER41020.
Use of the Stanford Synchrotron Radiation Lightsource, SLAC National Accelerator Laboratory, is supported by the U.S. Department of Energy, Office of Science, Office of Basic Energy Sciences under Contract No. DE-AC02-76SF00515.
The group from USP acknowledges support from FAPESP (grant 2020/04867-2) and CAPES.

\bibliography{bib/TechnicalProposal,bib/hpk_fbk_paper,bib/HGTD_TDR,bib/data_analysis, bib/others}
}

\end{document}